\documentclass[onecolumn,showpacs,amsmath,amssymb,12pt]{revtex4}
\usepackage{bm,amsmath,amssymb} \usepackage{graphicx}
\usepackage{amssymb}
\usepackage{epsfig}
\usepackage[english]{babel}
\usepackage{latexsym}
\usepackage{graphics}
\usepackage{epsfig}
\usepackage{graphicx}
\usepackage{dcolumn}
\usepackage{amsmath}

\begin{document}


\title{Understanding two-photon double ionization of helium from the perspective of the characteristic time of dynamic transitions}


\author{Fei Li$^{1,8}$, Facheng Jin$^{1,2}$, Yujun Yang$^{3}$, Jing Chen$^{4,5}$, Zong-Chao Yan$^{6,7}$, Xiaojun Liu$^{7}$ and Bingbing Wang$^{1,8}$}
\email{wbb@aphy.iphy.ac.cn}

\address{$^1$Laboratory of Optical Physics, Beijing National Laboratory for Condensed Matter Physics, Institute of Physics, Chinese Academy of Sciences, Beijing 100190, China}
\address{$^2$Faculty of Science, Xi'an Aeronautical University, Xi'an 710077,  China}
\address{$^3$Institute of Atomic and Molecular Physics, Jilin University, Changchun 130012, China}
\address{$^4$Institute of Applied Physics and Computational Mathematics, P. O. Box 8009, Beijing 100088, China}
\address{$^5$Department of Physics, College of Science, Shantou University,
Shantou, Guangdong 515063, People's Republic of China}
\address{$^6$Department of Physics, University of New Brunswick, Fredericton, New Brunswick, Canada E3B 5A3}

\address{$^7$State Key Laboratory of Magnetic Resonance and Atomic and Molecular Physics, Wuhan Institute of Physics and Mathematics, Chinese Academy of Sciences, Wuhan 430071, China}

\address{$^8$University of Chinese Academy of Sciences, Beijing 100049,  China}
\date{\today}

\begin{abstract}
By using the B-spline numerical method, we investigate a two-photon double-ionization (TPDI) process of helium in a high-frequency laser field with its frequency ranging from 1.6~a.u. to 3.0~a.u. and the pulse duration ranging from 75 to 160~attoseconds. We found that there exists a characteristic time $t_{c}$ for a TPDI process, such that the pattern of energy distribution of two ionized electrons presents a peak or two, depending respectively on whether the pulse duration is shorter or longer than $t_{c}$. Especially, as the pulse duration is larger than $t_c$, the TPDI spectrum shows a double-peak structure which is attributed to the fact that most of the electron-electron Coulomb interaction energy is acquired by single electron during their oscillation around the nucleus before the two electrons leave, and hence the double-peak structure cannot be identified as a signal of sequential ionization. Additionally, if the photon energy is less than the ionization energy of He$^{+}$, {\it i.e.,} the photon energy is in the so called nonsequential ionization region, $t_{c}$ is not a fixed value, and it increases as the photon energy decreases; while if the energy of a photon is greater than the ionization energy of He$^{+}$, {\it i.e.,} the photon energy is in the so called sequential ionization region, $t_{c}$ is fixed at about 105 attoseconds. We further found that, for a helium-like ion in its ground state, the characteristic time for the case of the photon energy larger than the ionization energy of the second electron has a key relation with the Coulomb interaction energy $\overline{V}_{12}$ between the two electrons, which can be expressed as $t_{c}\overline{V}_{12}=4.192$, a type of quantum mechanical uncertainty relation between time and energy. In addition, this relation can be attributed to the existence of a minimal evolution time from the ground state to a double ionization state with two electrons carrying different energies. These results may shed light on deeper understanding of many-electron quantum dynamical processes.
\end{abstract}

 \maketitle



\section{INTRODUCTION}
With the development of free-electron laser technology~\cite{Inagaki2008,Allaria2012,Gauthier2016}, it will soon become possible to explore experimentally multi-electron dynamics by ultrashort extreme ultraviolet light pulses, which is critically important for us to understand many phenomena, such as superconductivity, molecular structure, and chemical reaction~\cite{Hu2013}. Two-photon double ionization (TPDI) of helium is one of the simplest multi-electron dynamical processes that has been the subject of intensive studies in the past decade or so. Most studies have mainly focused on the cross sections~\cite{Kurka2010,Palacios2009A,Horner2009,Feist2008,Guan2008,Zhang2011}, recoil ion momentum distributions~\cite{Kurka2010,Rudenko2008}, nuclear recoil differential cross sections~\cite{Horner2010,Abdel-Naby2013,Jiang2014}, and the role of  electron correlations~\cite{Foumouo2008,Kurka2010,Zhang2011,Feist2009J,Pazourek2012,Jiang2017,Hu2017} and their probing and controlling~\cite{Bergues2012,Hu2013,Feist2009L,Waitz2017,Winney2017}. A thorough understanding of these dynamical processes for two correlated electrons can help us understand multi-electron dynamical processes in more complex systems.

The occurrence of a TPDI process in helium depends on the photon energy $\omega$. In order to ionize the two electrons of helium, the total energy of the two photons must be greater than the whole ionization energy of the atom, {\it i.e.,} $2\omega>I_{p}$, where $I_{p}\approx2.9$~a.u.. Due to the opposite spins of the two electrons in the ground state of helium, the two electrons can occupy the same spatial state. When an ultrashort intense laser pulse with the photon energy $\omega>I_{p}/2$ is applied to the helium atom, each of the two electrons is possibly ionized by absorbing one photon separately under the joint influence of the nucleus, the laser pulse, and the other electron during the process of TPDI. Owing to the same initial state for the two electrons, irrespective of the spin state of each electron, these electrons are subject to the same influence from the nucleus and the laser field during the TPDI process. Therefore, the difference or similarity of final energy distributions of the two ionized electrons is only caused by their mutual Coulomb interaction.

On the other hand, in order to explore the electron-electron correlations in the ground state of helium from the time-domain perspective, Foumouo
{\it et al.}~\cite{Foumouo2008} defined a characteristic timescale associated with the dielectronic interaction energy $E_{\rm int}$, the one that the two electrons can exchange during a TPDI process. In their paper, the timescale is estimated roughly as $2\pi/E_{\rm int}=140$~attosecond (asec) with $E_{\rm int}=2I_{p_2}-I_{p_1}-I_{p_2}=1.1$~a.u., where $I_{p_2}$ is the ionization energy of He$^+$ and $I_{p_1}$ the first ionization energy of helium. In this work, we try to answer the following questions. Since the two electrons in the ground state of helium are in the same situation, can they carry the same kinetic energy if they are ionized by absorbing two photons very quickly? If they can, then how fast they can be ionized in order to carry the same kinetic energy? To answer these questions, for a TPDI process of helium we define a characteristic time by sweeping the pulse duration to obtain the turning point for the energy distribution of the two ionized electrons changing from one peak ({\it i.e.}, the two electrons carry the same kinetic energy) to two peaks. Interestingly, we find that this characteristic time maintains a constant when the photon energy of the laser field is larger than the ionization threshold of He$^+$, whereas it increases as the frequency decreases when the photon energy is less than the ionization energy. By defining the Coulomb interaction energy as the expectation value of the Coulomb interaction potential
between the two electrons over the ground-state helium wave function, we establish a relation between the characteristic time and Coulomb interaction energy that is similar to the quantum time-energy uncertainty relation. We further find that such a relation holds also for helium-like ions.

\section{THEORETICAL METHOD}
We solve the time-dependent Schr\"{o}dinger equation to study the interaction of the atomic helium with an intense laser pulse by using the B-spline method~\cite{Shi2001}. We first solve the following energy eigenvalue problem variationally for a field-free helium-like system (atomic units are used throughout, unless otherwise stated)
\begin{equation}\label{equa1}
  H_{0}\Psi_{n}(\textbf{r}_{1},\textbf{r}_{2})=E_{n}\Psi_{n}(\textbf{r}_{1},\textbf{r}_{2}),
\end{equation}
where $H_{0}$ is the Hamiltonian of the system
\begin{equation}\label{equa2}
  H_{0}=-\frac{1}{2}\nabla_{1}^{2}-\frac{1}{2}\nabla_{2}^{2}-
  \frac{Z}{r_{1}}-\frac{Z}{r_{2}} +\frac{1}{|\textbf{r}_{1}-\textbf{r}_{2}|}\,,
\end{equation}
with $Z$ being the nuclear charge.
According to the variational principle, $\Psi_{n}(\textbf{r}_{1},\textbf{r}_{2})$ can be solved by selecting an appropriate trial wave function expanded in terms of a complete set
\begin{equation}\label{equa3}
  \Psi_{n}(\textbf{r}_{1},\textbf{r}_{2})=\sum_{\substack{i_{\alpha}, i_{\beta},\\ \ell_{\alpha}, \ell_{\beta}} }C_{i_{\alpha}, i_{\beta},\ell_{\alpha}, \ell_{\beta}}[1+(-1)^{S}P_{12}]B_{i_{\alpha}}^{k}(r_{1})B_{i_{\beta}}^{k}(r_{2}) \sum_{m_{\alpha}m_{\beta}}\langle \ell_{\alpha}m_{\alpha}\ell_{\beta}m_{\beta}|LM \rangle Y_{\ell_{\alpha}}^{m_{\alpha}}(\hat{\textbf{r}}_{1})Y_{\ell_{\beta}}^{m_{\beta}} (\hat{\textbf{r}}_{2}),
\end{equation}
where $P_{12}$ is the permutation operator between electrons 1 and 2, $B_{i_{\alpha}}^{k}(r_{1})$ and $B_{i_{\beta}}^{k}(r_{2})$ are two B-spline functions of order $k$, $\langle \ell_{\alpha}m_{\alpha}\ell_{\beta}m_{\beta}|LM \rangle$ is the Clebsch-Gordan coefficient, $Y_{\ell_{\alpha}}^{m_{\alpha}}(\hat{\textbf{r}}_{1})$ and $Y_{\ell_{\beta}}^{m_{\beta}} (\hat{\textbf{r}}_{2})$ are the two spherical harmonics, and $S$, $L$, and $M$ are, respectively, the total electron spin, the total angular momentum and its $z$-component. In our calculation, the order of B-spline functions is chosen to be $k=7$, the field-free eigenfunctions of helium are obtained in a radial box extending up to 60 a.u. with 102 B-spline functions in each radial coordinates, and the angular momentum pair $(\ell_{\alpha},\ell_{\beta})$ is up to (4,4).

In the length gauge and the electric dipole approximation, the time-dependent Schr\"{o}dinger equation reads
\begin{equation}\label{equa4}
  i\frac{\partial}{\partial t}\Phi(\textbf{r}_{1},\textbf{r}_{2},t)=[H_{0}+\textbf{E}(t)\cdot(\textbf{r}_{1}+\textbf{r}_{2})] \Phi(\textbf{r}_{1},\textbf{r}_{2},t),
\end{equation}
where $\textbf{E}(t)$ is the electric field of the laser pulse. The electric field can be expressed in the form
\begin{equation}\label{equa5}
  \textbf{E}(t)=E_{0}f(t)\textmd{cos}(\omega t)\hat{\textbf{e}}_{z},
\end{equation}
where $E_{0}$ is the electric-field amplitude, $f(t)=e^{-(2\ln2)t^2/\tau^{2}}$ is the Gaussian envelope of the laser pulse with the full width at half maximum (FWHM) $\tau$, $\omega$ is the frequency of the laser pulse, and $\hat{\textbf{e}}_{z}$ is the unit vector of the laser polarization direction. We should mention that the pulse duration $\tau$ in this work is defined by the FWHM of intensity, which corresponds to $\sqrt{2}/2$ times the pulse duration used by Feist {\it et al.}~\cite{Feist2009J} and $\sqrt{2}/4$ times the pulse duration used by
Palacios and Foumouo {\it et al.}~\cite{Palacios2009A,Foumouo2008}. The two-electron time-dependent wave function can be expanded in terms of the field-free atomic eigenfunctions
\begin{equation}\label{equa6}
  \Phi(\textbf{r}_{1},\textbf{r}_{2},t)=\sum_{n}a_{n}(t)e^{-iE_{n}t}\Psi_{n}(\textbf{r}_{1},\textbf{r}_{2}).
\end{equation}
Substituting Eq.~(\ref{equa6}) into Eq.~(\ref{equa4}), one can obtain a set of coupled differential equations, which can be solved by the Adams method~\cite{Adam_method}. Once the time-dependent wave function $\Phi(\textbf{r}_{1},\textbf{r}_{2},t)$ is determined, the  probability distribution
at time $t_f$ for the two ionized electrons escaped with momenta ${\bf k}_{1}$ and ${\bf k}_{2}$ is obtained according to
\begin{equation}\label{equa7}
  P({\bf k}_{1},{\bf k}_{2})=|\langle\psi_{\textbf{k}_{1},\textbf{k}_{2}}(\textbf{r}_{1},\textbf{r}_{2})
  |\Phi(\textbf{r}_{1},\textbf{r}_{2},t_{f})\rangle|^{2},
\end{equation}
where $\psi_{\textbf{k}_{1},\textbf{k}_{2}}(\textbf{r}_{1},\textbf{r}_{2})$ is the uncorrelated double continuum state, which can be constructed as a product of two independent-particle Coulomb wave functions~\cite{Peng2010}
\begin{equation}\label{equa8}
  \psi_{\textbf{k}_{1},\textbf{k}_{2}}(\textbf{r}_{1},\textbf{r}_{2})=\frac{1}{\sqrt{2}}
  [\phi_{\textbf{k}_{1}}^{(-)}(\textbf{r}_{1})\phi_{\textbf{k}_{2}}^{(-)}(\textbf{r}_{2})+
  \phi_{\textbf{k}_{2}}^{(-)}(\textbf{r}_{1})\phi_{\textbf{k}_{1}}^{(-)}(\textbf{r}_{2})].
\end{equation}
Therefore, the energy distribution of the two ionized electrons can be written as
\begin{equation}\label{equa9}
  P(E_{1},E_{2})=\int\int k_{1}k_{2}P({\bf k}_{1},{\bf k}_{2})\,d\hat{\textbf{k}}_{1}d\hat{\textbf{k}}_{2},
\end{equation}
where $E_{1}=k_{1}^{2}/2$ and $E_{2}=k_{2}^{2}/2$ are the energies of the two ionized electrons.

\section{NUMERICAL RESULTS}

We now consider a TPDI process of helium in an ultrashort laser field with FWHM being 75 asec. The energy distribution of the two escaped electrons at the end of simulation is shown in Fig.~\ref{fig1} for the peak intensity $I=\rm{1\times10^{14}\;W/cm^{2}}$ and laser frequency (a) 2.4~a.u., (b) 2.1~a.u., (c) 1.8~a.u., and (d) 1.6~a.u.. We found that the pattern of the energy distribution presents a single elliptical peak in Fig.~\ref{fig1}(a), where the lengths of semi-major and -minor axes are, respectively, $a\rm{=0.9161}$ and $b\rm{=0.4633}$ obtained by fitting the contour curve corresponding to the half maximum probability density. In addition, the center of the ellipse is at the point of the maximum probability density with its coordinates $(E_c,E_c)=(0.7043, 0.7043)$,  which is a little smaller than the expected value $E_s=(2\omega-I_p)/2$~\cite{ee}, where $I_p$ is the ionization energy of helium. When the frequency of the laser pulse decreases from 2.4~a.u. to 1.6~a.u., we found that the pattern of the energy distribution keeps the shape of single elliptical peak with the center of the ellipse decreasing along the straight line $E_{1}=E_{2}$ according to the energy conservation with the difference between the center coordinate value $E_c$ and the expected value $E_s$ shown in Table~\ref{table_1}. The lengths of the semi-major and -minor axes of the elliptical peak shown in Fig.1~(b) are changed slightly to $a\rm{=0.9415}$ and $b\rm{=0.3816}$, whereas the lengths of the two axes of the corresponding ellipses cannot be obtained for the cases of Figs.~\ref{fig1}(c) and \ref{fig1}(d), because the energy distribution becomes a part of the single elliptical peak. Furthermore, we investigate the effect of the laser intensity on the pattern of the energy distribution. We found that the pattern of the energy distribution keeps unchanged as the laser intensity decreases from $\rm{1\times10^{14}\;W/cm^{2}}$ to $\rm{1\times10^{13}\;W/cm^{2}}$ and $\rm{1\times10^{12}\;W/cm^{2}}$, where only the value of the corresponding probability density is reduced by two and four orders of magnitude respectively.

We now investigate a TPDI process of helium in a longer laser pulse of FWHM 160~asec. The energy distribution of the two escaped electrons is shown in Fig.~\ref{fig2} for the peak intensity $I=\rm{1\times10^{14}\;W/cm^{2}}$ and laser frequency (a) 2.4~a.u., (b) 2.1~a.u., (c) 1.8~a.u.,
and (d) 1.6~a.u.. We found that the pattern of the energy distribution now presents two elliptical peaks that are symmetrical about the $E_{1}=E_{2}$ line. In Fig.~\ref{fig2}(a), the center of the ellipse above the $E_{1}=E_{2}$ line is at the point of the maximum probability density with the coordinates (0.4128, 1.3533), and the lengths of semi-major and -minor axes are, respectively, $a\rm{=0.3215}$ and $b\rm{=0.1689}$ obtained by fitting a part of contour curve above the $E_{1}=E_{2}$ line corresponding to the half maximum of the probability density. When the frequency of the laser pulse decreases from 2.4~a.u. to 1.6~a.u., as shown in Figs.~\ref{fig2}(a)-(d), the pattern of the energy distribution keeps the shape of double elliptical peaks, and the center of the pattern moves along the line of $E_{1}=E_{2}$ towards a low energy region, which can also be confirmed by fitting the corresponding contour curve. The lengths of the elliptical axes become $a\rm{=0.5085}$ and $b\rm{=0.2004}$ for the case of $\omega=2.1$~a.u., while the lengthes of the two axes of the corresponding ellipses for the case of $\omega=1.8$~a.u. (Fig.~\ref{fig2}(c)) and $\omega=1.6$~a.u. (Fig.~\ref{fig2}(d)) cannot be obtained since the energy distribution becomes a part of the double elliptical peaks. When the peak intensity of the laser pulse decreases to $\rm{1\times10^{13}\;W/cm^{2}}$ and $\rm{1\times10^{12}\;W/cm^{2}}$, we also found that the pattern of the energy distribution remains the same for the corresponding laser frequency, but the value of the corresponding probability density is reduced by two and four orders of magnitude, respectively.

From above results, we may find that the TPDI spectrum changes from one peak into two peaks with the increase of the pulse duration no matter whether the photon energy of the laser is smaller or larger than the ionization energy of He$^+$ ion. Then we would ask the question: how does the two-peak structure form during the TPDI? To answer this question, we first study the two-electron radial density distribution and the conditional angular distribution~\cite{Feist2009L} in position space at different moments. Figure~\ref{fig3} shows the electron radial density distribution $\rho (r_1,r_2,t) = r_{1}^{2}r_{2}^{2}\int\int d\Omega_{1}d\Omega_{2}|\Phi(\textbf{r}_{1},\textbf{r}_{2},t)|^2$ for the sum of $S$ (except the 1$S$ state) and $D$ states under the same laser conditions as in Fig.~\ref{fig1}(a). In order to eliminate the obstruction from single ionization states and bound states~\cite{Scrinzi2008}, we only select the region of $r_{1}\geqslant 3$~a.u. and $r_{2}\geqslant3$~a.u., where the
boundary 3~a.u. is chosen based on the electron radial density distribution of the ground state of $\rm{He^{+}}$ ion. It is found that the wave packet is located in the space close to the nucleus at the peak of the laser pulse as shown in Fig.~\ref{fig3}(a). Then the wave packet spreads away from the nucleus and gradually presents a symmetrical band-like distribution about the $r_{1}=r_{2}$ line as shown in Figs.~\ref{fig3}(b-g). After the end of the laser pulse, the two electrons mainly move away from the nucleus equidistantly as shown in Figs.~\ref{fig3}(h-i). The evolution of the wave packet, as shown in Fig.~\ref{fig3}, indicates that the two electrons are always located in the vicinity of the nucleus under the joint influence of the laser field and the Coulomb interaction; in other words, the electrons have no chance to escape from the parent ion until the end of the laser pulse.

Figure~\ref{fig4} shows the conditional angular distribution under the same laser conditions as in Fig.~\ref{fig1}(a). The conditional angular distribution $P(\theta_{12},\theta_{1}=0^{\circ},t)$ is defined as the probability of TPDI as a function angle $\theta_{12}$ between the position vectors of the two electrons in the coplanar geometry, {\it i.e.,} the azimuthal angles $\phi_{1}=\phi_{2}=0^{\circ}$ or $180^{\circ}$, where the position vector of the first electron lies along the laser polarization direction $\theta_{1}=0^{\circ}$. The probability is obtained by integrating the radial variables $r_{1}$ and $r_{2}$ over the region of $r_{1}\geqslant 3$~a.u. and $r_{2}\geqslant3$~a.u.:  $P(\theta_{12},\theta_{1}=0^{\circ},t)=\int_{3}^{r_{m}}\int_{3}^{r_{m}}dr_{1}dr_{2}r_{1}^{2}r_{2}^{2} |\Phi(r_{1},r_{2},\theta_{1}=0^{\circ},\theta_{12}=\theta_{2},\phi_{1}=0^{\circ},\phi_{2}=0^{\circ},t)|^{2}$, where the upper bound of the integral is $r_m=60$~a.u.. We found that the two electrons are most likely to distribute on both sides of the nucleus at the peak laser pulse as shown in Fig.~\ref{fig4}(a). Then two electrons oscillate around two sides of the nucleus driven by the laser field as shown in Figs.~\ref{fig4}(b-g). Finally, the two electrons keep on both sides of the nucleus after the time $t=2$ o.c. as shown in Figs.~\ref{fig4}(h-i), indicating that the two electrons are emitted mainly back-to-back after the end of the laser pulse. This result can be attributed to the fact that the Coulomb interaction energy between the two electrons can still be felt as shown in Fig.~\ref{fig3}(h) after the end of the laser pulse; hence the side-by-side emission is suppressed by the electron correlation. By analyzing the TPDI process for different times under such an ultrashort laser pulse as shown in Figs.~\ref{fig3} and \ref{fig4}, we may see that the single peak structure of the energy distribution shown in Fig.~\ref{fig1}(a) can be understood as follows: Since the duration of the laser pulse is very short, there is no enough time for the two electrons to share the Coulomb interaction energy during the laser pulse, and majority part of this interaction energy is shared almost equally by the two electrons when they are emitted back-to-back from both sides of the nucleus equidistantly after the end of the laser pulse.

We then look at the two-electron radial density distribution and the conditional angular distribution for the longer laser pulse. Figure~\ref{fig5} shows the radial density distribution of $S$ (except the 1$S$ state) and $D$ states of helium at different moments under the same laser conditions
as in Fig.~\ref{fig2}(a). We see that the wave packet of the two electrons is also located in the space close to the nucleus at the peak laser pulse as shown in Fig.~\ref{fig5}(a). Then the wave packet spreads away from the nucleus, and at the same time presents a band-like structure at $t=(n+1/2)/2$ o.c. with $n$=0, 1, 2, 3 and a finger-like structure at $t=n/2$ o.c. with $n$=1, 2, 3, 4 as shown in Figs.~\ref{fig5}(b-i). At last, the distribution remains the finger-like structure after $t=2$ o.c. as shown in Figs.~\ref{fig5}(j-l), indicating that one electron is farther away from the nucleus than the other one. Therefore, the electron that is farther away from the nucleus carries higher energy, which implies that majority part of the two-electron Coulomb interaction energy is acquired by this electron before the end of laser pulse. To understand the mechanism more clearly, Fig.~\ref{fig6} shows the conditional angular distribution under the same laser conditions as in Fig.~\ref{fig2}(a). Comparing with Fig.~\ref{fig4}, we found that the two electrons also oscillate between the two sides of the nucleus in the laser field before $t=2$ o.c.. However, the probability distribution of the two electrons gradually reaches a balance between the two electrons being on the two sides of the nucleus and being on one side after $t=2$ o.c., as shown in Figs.~\ref{fig6}(j-l), implying that the contribution is almost equal for both the back-to-back emission and the side-by-side emission. This result shows that the two electrons leave with weak electron correlation as shown in Fig. 5(l). As a result, the energy distribution for the case of the longer laser pulse presents two elliptical peaks as shown in Fig.~\ref{fig2}(a), which can be attributed to the fact that almost all the Coulomb interaction energy is acquired by one electron before the end of the laser pulse and the correlation between the two electrons is weak at the same time.

Finally, we investigate the TPDI process of helium for different pulse durations. Figure~\ref{fig7} is the energy distribution of the two ionized electrons with the intensity of $\rm{1\times10^{14}\;W/cm^{2}}$, the frequency $\omega=2.4$~a.u., and the FWHM (a) 80~asec, (b) 100~asec, (c) 120~asec, and (d) 140~asec. We can see that the energy distribution changes from a single elliptical peak to two elliptical peaks when the FWHM of the laser pulse increases from 80~asec to 140~asec, and the turning point for this change is around 105~asec, {\it i.e.}, 4.339~a.u..

As we have known from the above analysis, the existence of the double-peak structure in the energy distribution
implies that the majority of the Coulomb interaction energy between the two electrons is acquired by one electron before the end of the laser pulse, while the single-peak structure implies that the Coulomb interaction energy is shared equally by both electrons after the end of laser pulse. In other words, we may define the characteristic time as the minimum time that the Coulomb interaction energy can be carried by one electron in a TPDI process, which simply corresponds to the turning point of the energy distribution from the single peak to the double peaks. In order to identify the value of the characteristic time for a TPDI process, we define the ratio $\eta=P_{\rm mid}/P_{\rm max}$, where $P_{\rm max}$ is the maximum probability of TPDI in energy space with its coordinates $(E_{1}(P_{\rm max}),E_{2}(P_{\rm max}))$, and $P_{\rm mid}$ is the probability of TPDI at the intersection of the lines $E_{1}+E_{2}=E_{\rm tot}$ with $E_{\rm tot}=E_{1}(P_{\rm max})+E_{2}(P_{\rm max})$ and $E_1=E_2$. According to this definition, $\eta=1$ means that the energy distribution has one peak while $0<\eta<1$ means that the energy distribution has two peaks. Therefore, the pulse duration for the turning point of $\eta$ to be changed from one to less than one can be identified as the characteristic time of the atom-laser system. Figure~\ref{fig8} shows the ratio $\eta$ as a function of the pulse duration with the laser intensity of $\rm{1\times10^{14}\;W/cm^{2}}$ at different frequencies for the case of helium (a) and Li$^+$ (b). As shown in Fig.~\ref{fig8}(a), one may find that the curves may be classified into two types: When the laser frequency is greater than the ionization energy of He$^+$, all the curves are consistent with each other and have the same turning point at $t=105$~asec; on the other hand, when the laser frequency is less than the ionization energy of He$^+$, the curves are no longer consistent with each other and thus the characteristic time increases as the frequency decreases. Similarly, Fig.~\ref{fig8}(b) shows that the characteristic time for the case of Li$^+$ is about 65~asec. These results clearly show that the characteristic time tends to a minimum limit as the laser photon energy increases from smaller  to larger than the ionization threshold for the second electron.

From the above analysis we see that, when the pulse duration is less than the characteristic time, the Coulomb interaction energy of the ground state is almost equally shared by the two electrons, which results in the single-peak structure in the energy distribution; when the pulse duration is greater than the characteristic time, most of the Coulomb interaction energy is carried by one electron before the two electrons are ionized, which results in the two-peak structure in the energy distribution. Clearly, the Coulomb interaction between the two electrons plays a critical role on the energy distribution and it may thus be related to the characteristic time. Different from the dielectronic interaction energy defined in Ref.~\cite{Foumouo2008}, we define the Coulomb interaction energy as the expectation value of the Coulomb interaction potential between the two electrons over the ground-state wave function of helium, {\it i.e.}, $\overline{V}_{12}=<\Psi_{1S}|1/r_{12}|\Psi_{1S}>$, which is $\overline{V}_{12}=0.947$~a.u.. We believe that our definition of the Coulomb interaction energy carries the correlation information in the real helium atom, in comparison to the estimation in Ref.~\cite{Foumouo2008}. In order to find out the relationship between the Coulomb interaction energy and the characteristic time in a TPDI process, we change the nuclear charge of the helium-like ion and calculate the corresponding Coulomb interaction energy and the characteristic time. As the nuclear charge increases, the two electrons are more localized in a smaller space, resulting in an increase of the Coulomb interaction energy. Our numerical results are presented in Table~\ref{table_2} and the curve of the characteristic time as a function of the Coulomb interaction energy is plotted in Fig.~\ref{fig9}. Surprisingly, we found the following relation between the characteristic time and the Coulomb interaction energy
\begin{equation}\label{equa7}
  t_{c}\overline{V}_{12}=4.192,
\end{equation}
where the relative error is less than $3\%$, as shown in Table~\ref{table_2}.

We can see that the relation between the characteristic time and the Coulomb interaction energy is quite similar to the quantum mechanical uncertainty relation between time and energy, although we still need to figure out where the parameter 4.192 comes from. For the above relation, we have demonstrated that the two-electron Coulomb interaction energy determines the characteristic time for a TPDI dynamic process, which can be understood as the minimal evolution time from the ground state to the state that the Coulomb interaction energy is acquired by one electron according to the concept of the quantum speed limits~\cite{Mandelstam1945, Pires2016, Okuyama, Shanahan}. More importantly, we should mention here that this characteristic time-Coulomb interaction energy equation is a very general relationship for the photon energy of the laser pulse larger than the ionization energy of the second electron in TPDI, which provides an example that the time-energy uncertainty relation has more general applications in dynamic processes, although the constant is not $1/2$.

\section{CONCLUSIONS}

In conclusion, we have investigated the TPDI process of helium subjected to a high frequency ultrashort laser pulse. We find that there exists a characteristic time which may identify different allocation of the Coulomb interaction energy between the two electrons for the TPDI process: If the pulse duration is shorter than this characteristic time, then the Coulomb interaction is equally shared by the two electrons when they are released,
while if the pulse duration is longer than the characteristic time, then most of the Coulomb interaction is obtained by one electron before the two electron are released. In order to find the relation between the characteristic time and the Coulomb interaction energy of the two electrons in a ground state, we varied the nuclear charge of the helium-like ion and found the relation between the characteristic time for the case of the photon energy larger than the ionization energy of the second electron and the Coulomb interaction energy $t_{c}\overline{V}_{12}=4.192$, which is similar to the quantum uncertainty relation between time and energy. This work provides an example for showing that the interaction energy between particles and the minimum time that this energy is finished allocating has a simple relation under the same dynamic mechanism. This may be treated as an extension of the quantum uncertainty relation, and we hope more situations may be found in the future.

\section*{ACKNOWLEDGMENTS}

This work was supported by the National Natural Science Foundation of China under Grant Nos. 11425414, 11474348, 11774129 and 11774411, and National Key Research and Development Program (No. 2016YFA0401100).
ZCY was supported by NSERC of Canada and by the CAS/SAFEA International Partnership Program for Creative Research Teams.
We thank all the members of SFAMP club for helpful discussions.


\newpage{
\begin{figure}
\includegraphics[width=9 cm]{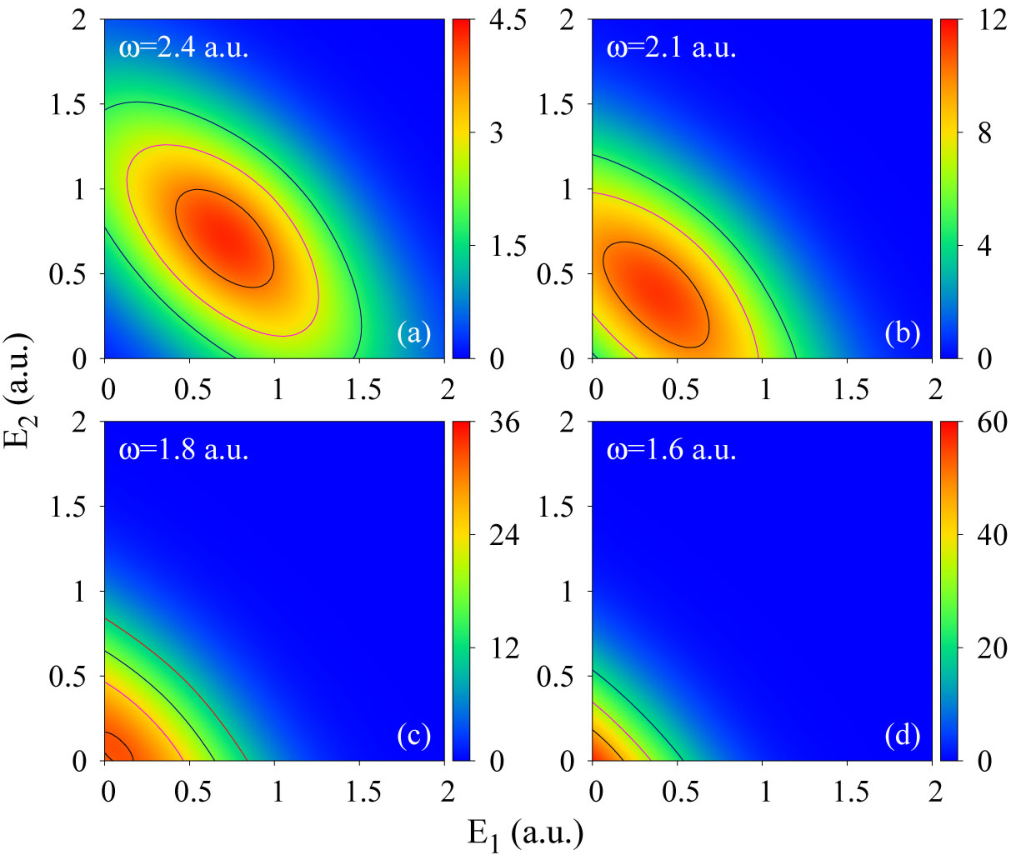}
\caption{(Color online) Energy distribution of two escaped electrons. The laser pulse has a Gaussian envelope around the peak intensity of $\rm{1\times10^{14}\,W/cm^{2}}$ and a time duration (FWHM) of 75 asec. The central photon energies are 2.4~a.u., 2.1~a.u., 1.8~a.u.,
and 1.6~a.u., respectively. The color bars are in units of $10^{-7}$. The curves represent contour lines.}
\label{fig1}
\end{figure}}

\begin{figure}
\includegraphics[width=9 cm]{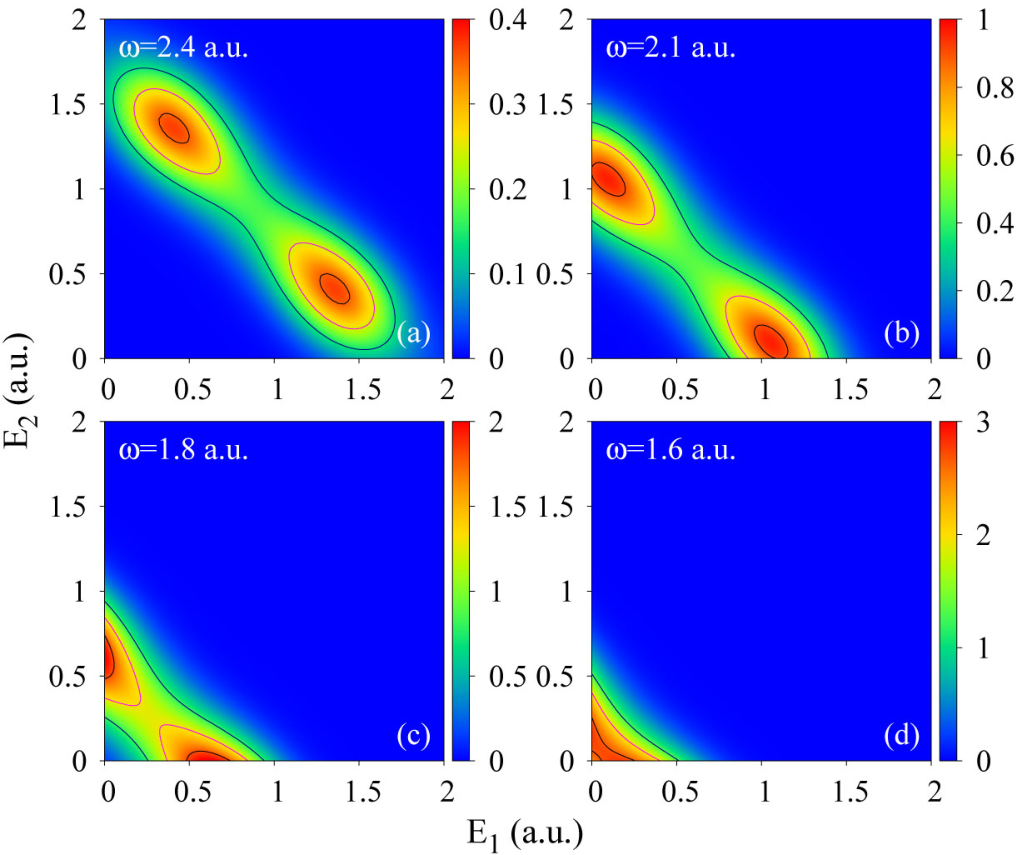}
\caption{(Color online) Energy distribution of two escaped electrons. The laser pulse has a Gaussian envelope around the peak intensity of $\rm{1\times10^{14}\,W/cm^{2}}$ and a time duration (FWHM) of 160 asec. The central photon energies are 2.4~a.u., 2.1~a.u., 1.8~a.u., and 1.6~a.u., respectively. The color bars are in units of $10^{-5}$. The curves represent contour lines.}
\label{fig2}
\end{figure}

\begin{figure}
\includegraphics[width=12 cm]{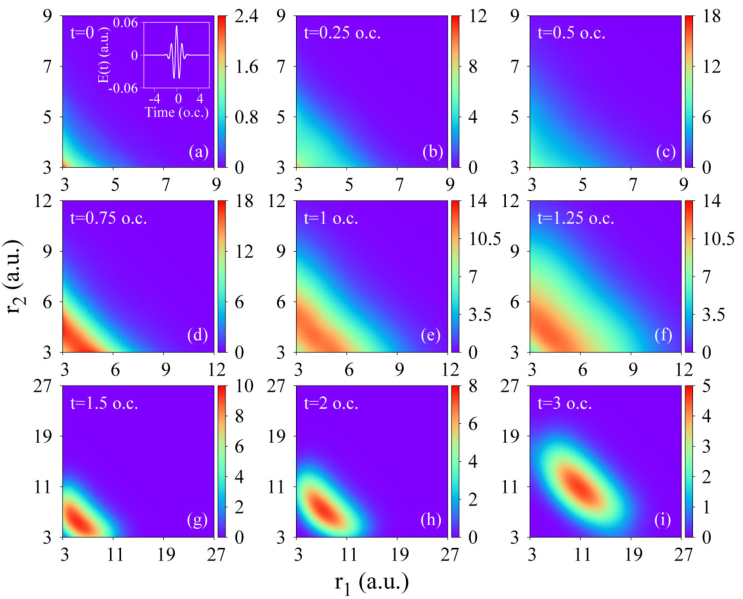}
\caption{(Color online) Radial density $\rho (r_1,r_2,t) = r_{1}^{2}r_{2}^{2}\int\int d\Omega_{1}d\Omega_{2}|\Phi(\textbf{r}_{1},\textbf{r}_{2},t)|^2$ at different times. The laser parameters are the same as Fig.~\ref{fig1}(a). The color bars are in units of $10^{-9}$. The inset in (a) shows the electric field of the laser pulse.}
\label{fig3}
\end{figure}

\begin{figure}
\includegraphics[width=12 cm]{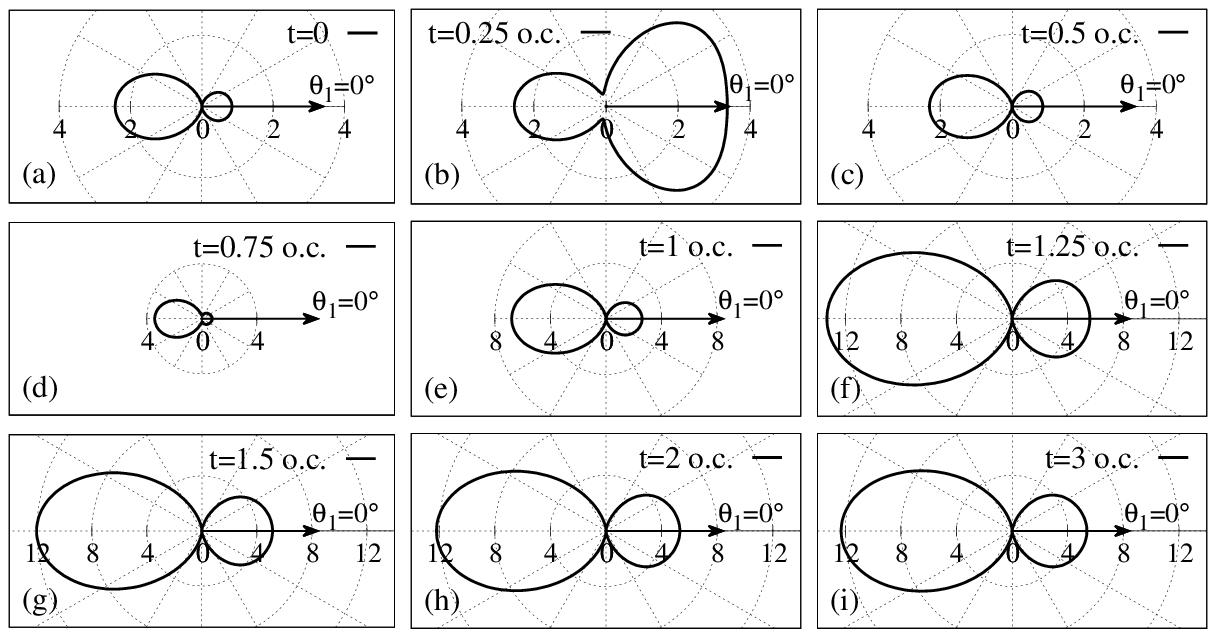}
\caption{Conditional angular distribution $P(\theta_{12},\theta_{1}=0^{\circ},t)=\int_{3}^{r_{m}}\int_{3}^{r_{m}}dr_{1}dr_{2}r_{1}^{2}r_{2}^{2} |\Phi(r_{1},r_{2},\theta_{1}=0^{\circ},\theta_{12}=\theta_{2},\phi_{1}=0^{\circ},\phi_{2}=0^{\circ},t)|^{2}$ of two electrons in position space at different times. The laser parameters are the same as Fig.~\ref{fig1}(a). In units of $10^{-10}$.}
\label{fig4}
\end{figure}

\begin{figure}
\includegraphics[width=12 cm]{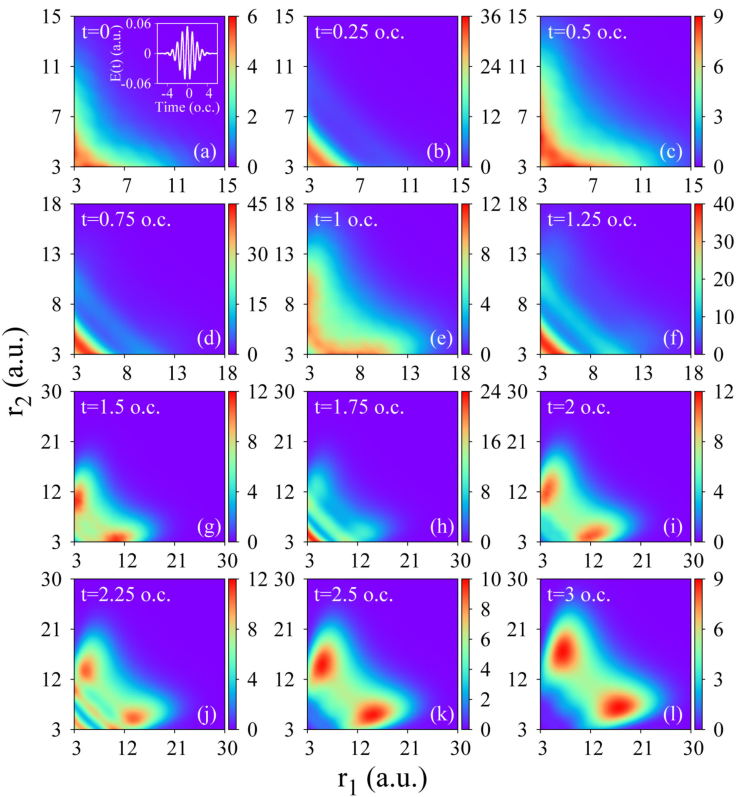}
\caption{(Color online). Radial density $\rho (r_1,r_2,t)=r_{1}^{2}r_{2}^{2}\int\int d\Omega_{1}d\Omega_{2}|\Phi(\textbf{r}_{1},\textbf{r}_{2},t)|^2$ at different times. The laser parameters are the same as Fig.~\ref{fig2}(a). The color bars are in units of $10^{-9}$. The inset in (a) shows the electric field of the laser pulse.}
\label{fig5}
\end{figure}

\begin{figure}
\includegraphics[width=12 cm]{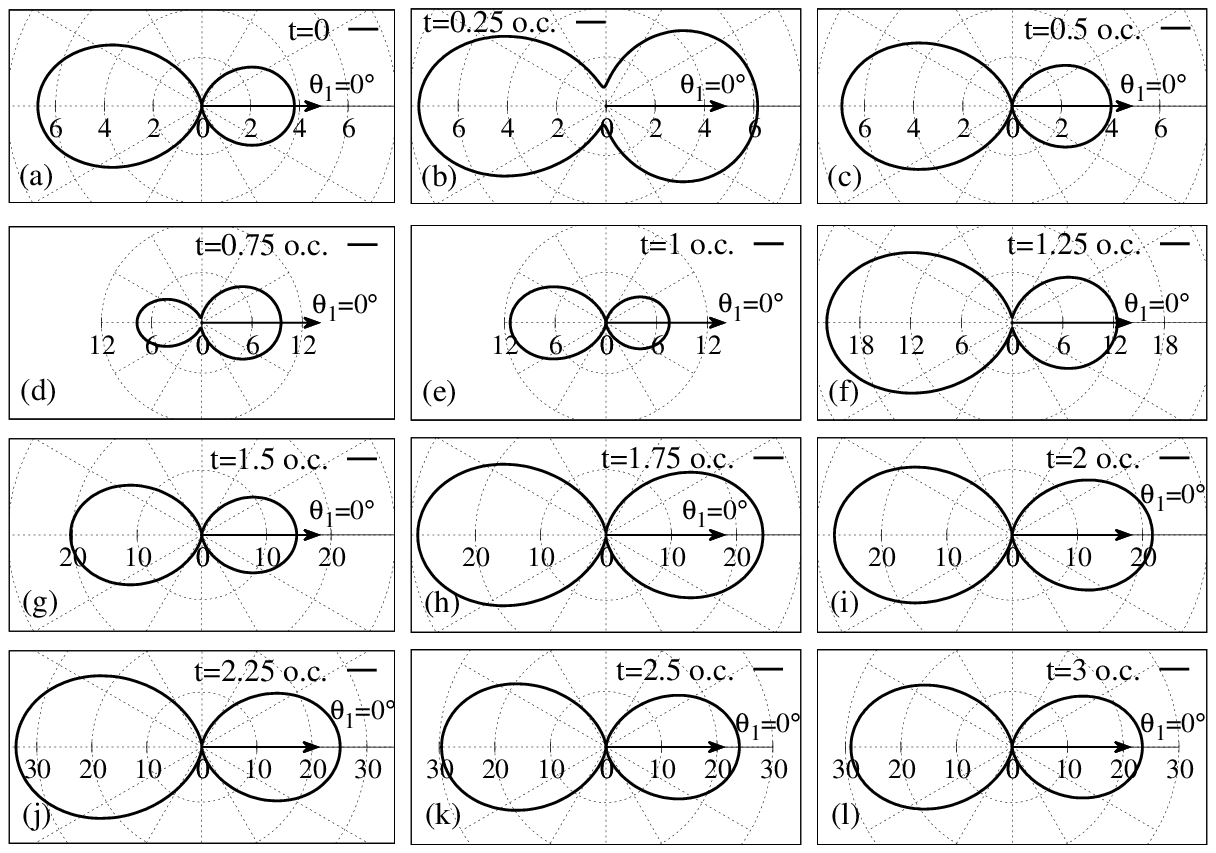}
\caption{Conditional angular distribution $P(\theta_{12},\theta_{1}=0^{\circ},t)=\int_{3}^{r_{m}}\int_{3}^{r_{m}}dr_{1}dr_{2}r_{1}^{2}r_{2}^{2} |\Phi(r_{1},r_{2},\theta_{1}=0^{\circ},\theta_{12}=\theta_{2},\phi_{1}=0^{\circ},\phi_{2}=0^{\circ},t)|^{2}$ of two electrons in position space at different times. The laser parameters are the same as Fig.~\ref{fig2}(a). In units of $10^{-9}$.}
\label{fig6}
\end{figure}

\begin{figure}
\includegraphics[width=8 cm]{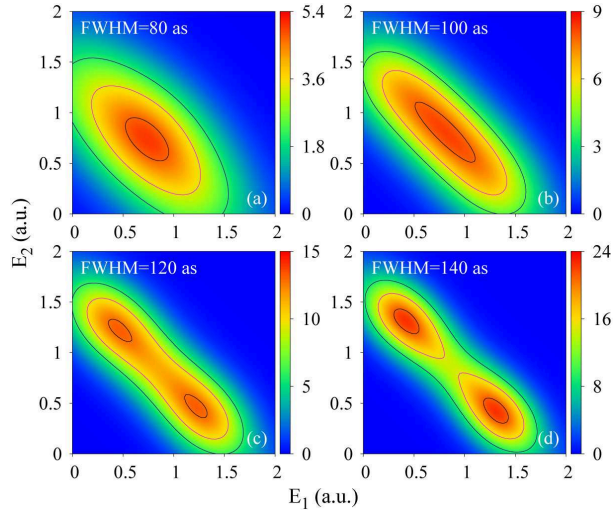}
\caption{(Color online) Energy distribution of two escaped electrons. The laser pulse has a Gaussian envelope around the peak intensity of $\rm{1\times10^{14}}\,W/cm^{2}$ and central photon energy of 2.4~a.u.. The time duration (FWHM) is 80~asec, 100~asec, 120~asec, and 140~asec, respectively. The color bars are in units of $10^{-7}$. The curves represent contour lines.}
\label{fig7}
\end{figure}

\begin{figure}
\includegraphics[width=8 cm]{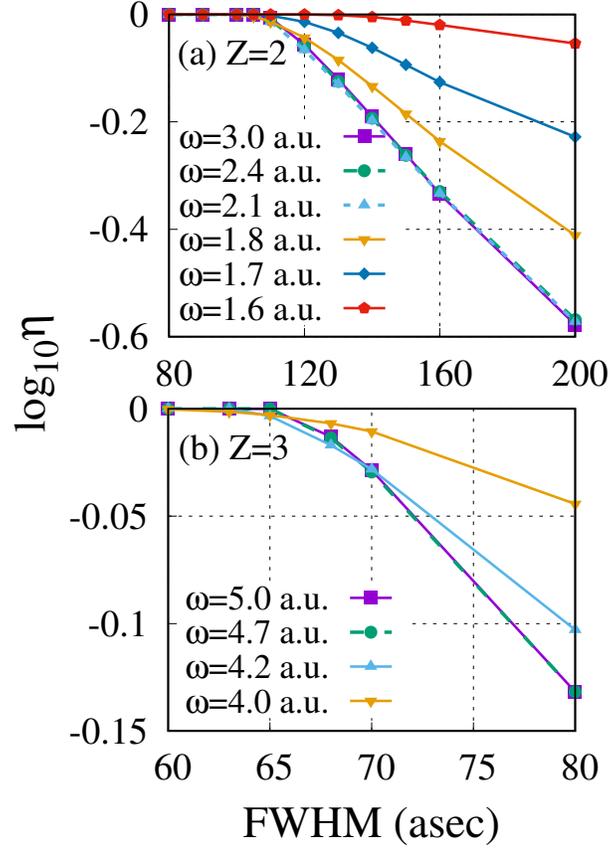}
\caption{The ratio $\log_{10}\eta$ varies with the FWHM of laser pusle. The laser pulse has a Gaussian envelope around the peak intensity of $\rm{1\times10^{14}}\,W/cm^{2}$. The central photon energy is 3.0~a.u., 2.4~a.u., 2.1~a.u., 1.8~a.u., 1.7~a.u., 1.6~a.u. for $Z=2$ (a); 5.0~a.u., 4.7~a.u., 4.2~a.u., 4.0~a.u. for $Z=3$ (b).}
\label{fig8}
\end{figure}

\begin{figure}
\includegraphics[width=9 cm]{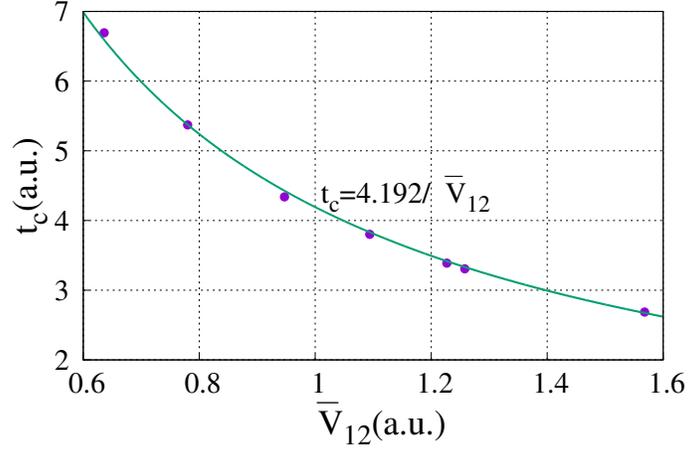}
\caption{The characteristic time of TPDI varies with the Coulomb interaction energy for different ground-state helium-like ions (solid circles). The laser pulse has a Gaussian envelope around the peak intensity of $\rm{1\times10^{14}}\,W/cm^{2}$. The solid line is the function $t_{c}=4.192/\overline{V}_{12}$.}
\label{fig9}
\end{figure}
\begin{table}[!tb]
\centering
\setlength{\abovecaptionskip}{10pt}
\setlength{\belowcaptionskip}{10pt}
\caption{\linespread{1.5}\selectfont The photon energy $\omega$, the center coordinate value $E_{c}$, its predicted value $E_s=(2\omega-I_{p})/2$, and energy difference $\Delta E=|(2\omega-I_{p})/2-E_{c}|$. The laser pulse peak intensity and pulse duration are $I=\rm{1\times10^{14}\,W/cm^{2}}$ and 75 asec, respectively. In atomic units.}\label{table_1}
\begin{tabular}{cccc}
  \hline
  \hline
  $\hspace{3mm}\omega\hspace{3mm}$ & $E_{c}$  & $(2\omega-I_{p})/2$ & $\Delta E$ \\
  \hline
  3        & 1.3479 & 1.5485 & 0.2006 \\
  2.4      & 0.7043 & 0.9485 & 0.2442 \\
  2.3      & 0.5978 & 0.8485 & 0.2507 \\
  2.2      & 0.4907 & 0.7485 & 0.2578 \\
  2.1      & 0.3817 & 0.6485 & 0.2668 \\
  \hline
  \hline
\end{tabular}
\end{table}

\begin{table}[!tb]
\centering
\setlength{\abovecaptionskip}{10pt}
\setlength{\belowcaptionskip}{10pt}
\caption{\linespread{1.5}\selectfont The nuclear charge $Z$, the two-electron Coulomb interaction energy $\overline{V}_{12}$, the first ionization potential $I_{p_1}$, the second ionization potential $I_{p_2}$, and the characteristic time $t_{c}$ obtained by the time-dependent Schr\"{o}dinger equation, where $\Delta=\Big|\frac{4.192-t_{c}\overline{V}_{12}}{t_{c}\overline{V}_{12}}\Big|$. In atomic units.}\label{table_2}
\begin{tabular}{cccccc}
  \hline
  \hline
  Z & $\overline{V}_{12}$ & $I_{p_1}$ & $I_{p_2}$ & $t_{c}$  &  $\Delta$ \\
  \hline
  1.5    & 0.636 & 0.340 & 1.125 & 6.694  & 0.0154  \\
  1.732  & 0.780 & 0.570 & 1.5   & 5.372  & 0.0005  \\
  2      & 0.947 & 0.903 & 2     & 4.339  & 0.0202  \\
  2.236  & 1.094 & 1.255 & 2.5   & 3.802  & 0.0079  \\
  2.45   & 1.227 & 1.624 & 3     & 3.388  & 0.0083  \\
  2.5    & 1.258 & 1.716 & 3.125 & 3.306  & 0.0080  \\
  3      & 1.568 & 2.780 & 4.5   & 2.686  & 0.0046  \\
  \hline
  \hline
\end{tabular}
\end{table}

\end{document}